\documentclass[10pt,letterpaper,twocolumn]{article} 

\usepackage{ol2}
\usepackage[draft]{hyperref}
\usepackage{amsmath}

\newcommand{\de}{\partial}
\newcommand{\sech}[1]{\textrm{sech}\left(  #1\right)}
\newcommand{\eq}[2]{\begin{equation} \label{#1} #2 \end{equation}}

\newcommand{\etal}{{\em et al.}}

\begin{document}

\twocolumn[ 

\title{Localized frequency comb and formation of embedded solitons in silicon-based slot waveguides}

\author{Samudra Roy and Fabio Biancalana}

\affiliation{Max Planck Institute for the Science of Light, G\"{u}nther-Scharowsky-Str. 1, Bau 26, 91058 Erlangen, Germany}

\date{\today}

\begin{abstract}
We explore the possibility to excite the so-called embedded solitons in specially designed slot waveguides based on silicon and silica or silicon nanocrystals. This requires the excitation of the structure with quasi-TM polarized pulses -- for which Raman effect is absent -- and at a specific infrared wavelength for which only the second- and fourth-order group velocity coefficients are non-vanishing. Pulses launched in these conditions will generate a spectrally localized continuum, associated to a frequency comb coming from the spectral interference of many embedded solitons.
\end{abstract}

    \pacs{060.5530
    , 190.4370
    , 230.4320
    }
]

In 1993, H\"o\"ok and Karlsson \cite{hook93} demonstrated theoretically the existence of a family of solitary waves, solution of the equation
\eq{eq1}{i\de_{z}A-\frac{1}{2}\beta_{2}\de_{t}^{2}A+\frac{1}{24}\beta_{4}\de_{t}^{4}A+\gamma|A|^{2}A=0,} where $A(z,t)$ is the electric field envelope, $\beta_{2,4}$ are respectively the second- and fourth-order group velocity dispersion (GVD) coefficients of the waveguide calculated at a reference frequency $\omega_{0}$ (taken to be the central frequency of the input pulse), and $\gamma$ is the nonlinear coefficient of the fundamental fiber mode. Eq. (\ref{eq1}) is valid when $\omega_{0}$ is located at a local maximum (or minimum) of the GVD, where $\beta_{3}$ vanished identically. The properties of such peculiar solitary waves -- sometimes called {\em embedded solitons} (ESs) due to their coexistence with linear waves \cite{kivsharbook} -- have been since then studied in detail. It was found that ESs exist only for $\beta_{2,4}<0$, can have or not oscillating decaying tails, can form symmetric (in-phase) and antisymmetric (out-of-phase) bound states, and that only the single peak solitons can be stable \cite{karlsson94,akhmediev94,buryak95,piche96}.

In principle, it would be possible to design specific solid-core photonic crystal fibers (PCFs) \cite{russell} with a quartic profile of the GVD as required by Eq. (\ref{eq1}), by using for instance the dispersion-flattened fibers reported in \cite{reeves}, or the nano-bore fibers reported in \cite{nanobore}. However, the perturbation induced by the Raman effect would completely destroy the ESs. Indeed, from the very beginning of the propagation, their central frequency would continuously translate due to the intrapulse Raman self-frequency shift \cite{mitschke}, thus moving all solitons away from the point at which $\beta_{3}=0$, which immediately breaks the validity of Eq. (\ref{eq1}). This is mainly the reason why the above theoretical works have been considered only as a mathematical curiosity, only marginally discussed in textbooks \cite{agrawalbook,kivsharbook}, and largely forgotten during the past two decades.

However, the recent advent of silicon photonics \cite{pavesi,jalali,lipson,linreview} has opened many interesting possibilities in nonlinear optics, mainly due to its potential applications in the spectral region extending from the near- to the mid-infrared. The high refractive index of silicon, combined with the silicon-on-insulator (SOI) technology, allows for a tight confinement of the optical modes and a consequent increase of the nonlinear coefficient of the waveguide, enabling efficient nonlinear optical interactions at low power levels and in relatively short length.

In this Letter we propose (for the first time to our knowledge) a feasible way to observe and use the generation of ESs in specially designed silicon-based waveguides.

\begin{figure}[htb]
\centering \includegraphics[width=0.5\textwidth]{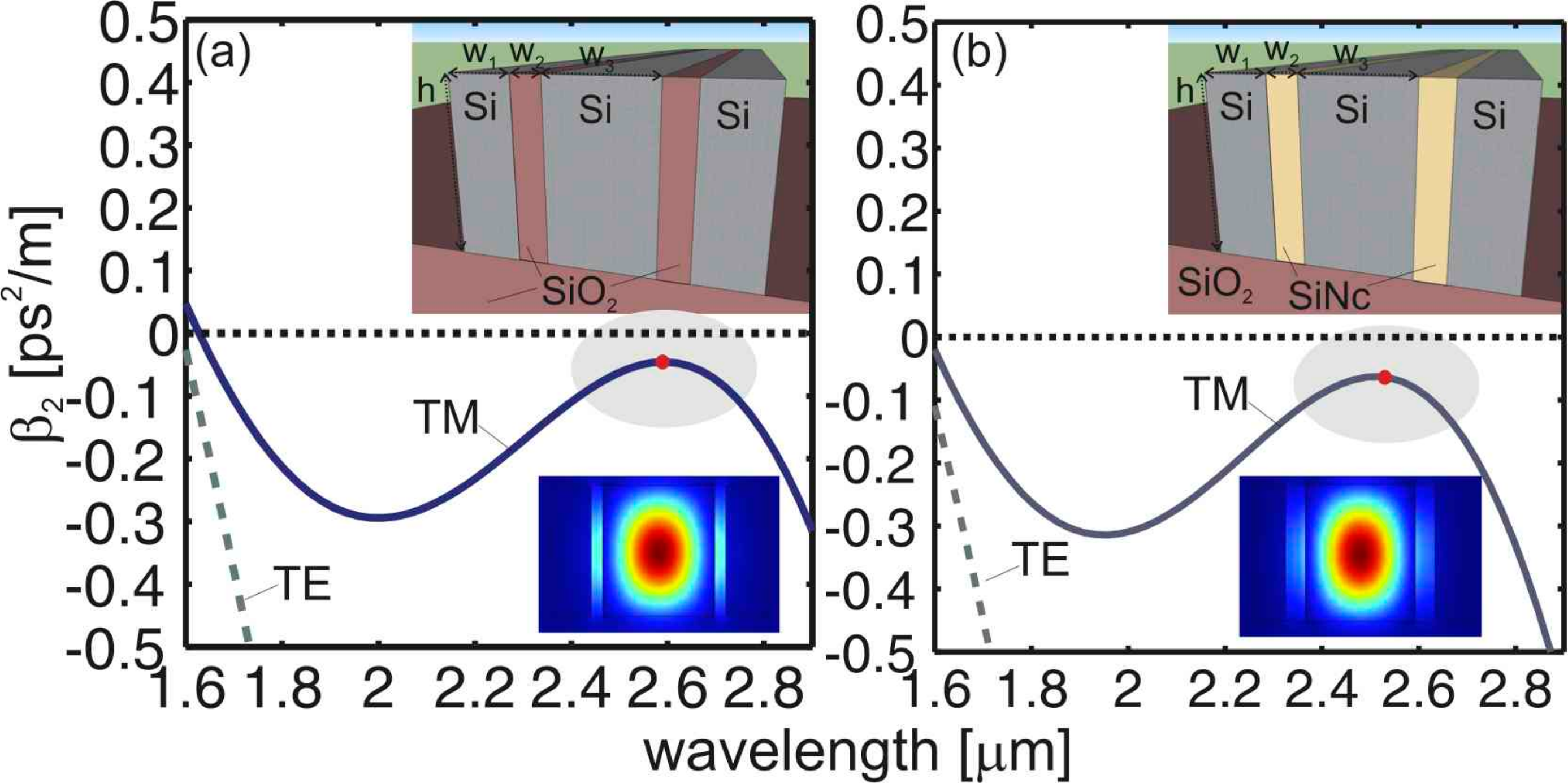}
  \caption{\small{(Color online) The silicon-based structures (both on a SiO$_{2}$ substrate) studied in this Letter. (a) GVD of quasi-TE (dashed line) and -TM (solid line) modes of the Si-SiO$_{2}$ slot waveguide. Inset below: TM-mode profile at $\lambda_{0}=2592$ nm. Inset above: geometry and materials of the structure. Parameters are: $h=700$ nm, $w_{1}=220$ nm, $w_{2}=60$ nm, $w_{3}=600$ nm. (b) GVD of quasi-TE (dashed line) and -TM (solid line) modes of the Si-SiNc slot waveguide. Inset below: TM-mode profile at $\lambda_{0}=2511$ nm. Inset above: geometry and materials of the structure. Parameters are: $h=700$ nm, $w_{1}=190$ nm, $w_{2}=80$ nm, $w_{3}=570$ nm. In both figures, the red dot indicates the wavelength ($\lambda_{0}$) at which $\beta_{3}=0$, and the gray area is the region where the quartic approximation of the GVD, and therefore the use of Eq. (\ref{eq2}), is valid.}}
  \label{fig1}
\end{figure}

The two structures that we propose -- both of which are built on a silica substrate -- have been inspired by Refs. \cite{zhang,zhang2}  and are depicted in Fig. \ref{fig1}(a,b). In Fig. \ref{fig1}(a) we show the structural parameters of our silicon-silica (Si-SiO$_{2}$) waveguide, where the GVD for quasi-TE and -TM modes and mode profile for quasi-TM mode are also shown. In Fig. \ref{fig1}(b) we show the same information as in Fig. \ref{fig1}(a), but for our silicon-silicon nanocrystals (Si-SiNc) waveguide. The gray-shaded circles, centered at the wavelengths $\lambda_{0}$ at which $\beta_{3}(\lambda_{0})=0$ (which from now will be referred as {\em quartic point}), are the regions around which the GVD can be approximated by including only $\beta_{2}$ and $\beta_{4}$ in the dispersion. Note that, in those regions, the dispersion is anomalous ($\beta_{2}<0$) and the GVD curvature is negative ($\beta_{4}<0$). As described above, these are exactly the necessary conditions for the possible existence of localized ESs, and the parameters of the structure must be carefully designed in order to achieve the correct sign and curvature of the GVD at the desired wavelength.

Apart from the shape of the GVD, one must consider the impact of the two-photon absorption (TPA) generated by the free carriers around $\lambda_{0}$. As it was demonstrated in numerous experimental works (see e.g. Refs. \cite{bristow,liu,jalali2,jalali3}), the impact of TPA is greatly reduced if $\lambda_{0}>2.2$ $\mu$m, away from the two-photon band edge, where the three-photon absorption is also insignificant \cite{jalali2,liu,jalali3}.

The final crucial property for the present work is that, for an SOI waveguide fabricated along the [$\bar{1}10$] direction on the [110]$\times$[001] surface, stimulated Raman scattering (SRS) cannot occur when an input pulse excites the quasi-TM mode of the waveguide \cite{lin,linreview}. Thus, for pulses launched close to $\lambda_{0}=2\pi c/\omega_{0}$, the nonlinearity is dominated by the Kerr effect, and is not affected by the carrier dynamics, and the dimensionless propagation equation can be thus written as
\eq{eq2}{i\de_{\xi}\psi+\frac{1}{2}\de_{\tau}^{2}\psi-|\delta_{4}|\de_{\tau}^{4}\psi+\left(1+is\de_{\tau}\right)|\psi|^{2}\psi=0,}
where we have performed the typical scalings $A\equiv\sqrt{P_{0}}\psi$, $t\equiv t_{0}\tau$, $z\equiv z_{0}\xi$, and where $z_{0}=t_{0}^{2}/|\beta_{2}|$, $\delta_{4}\equiv\beta_{4}/(24|\beta_{2}|t_{0}^{2})$, $P_{0}\equiv(\gamma z_{0})^{-1}$, $\gamma$ is the nonlinear coefficient of the structure calculated with a mode solver (mainly due to the field inside the central silicon part), and $s\equiv(\omega_{0}t_{0})^{-1}$, with $t_{0}$ being the input pulse duration \cite{agrawalbook}.


In Eq. (\ref{eq2}), it was very important to take into account the shock operator that multiplies the Kerr nonlinear term. Such operator considerably influences the dynamics in those situations where the Raman effect is absent, as in the present case. This constitutes the main difference between Eq. (\ref{eq1}) and Eq. (\ref{eq2}), the latter having now a direct physical interpretation in the systems studied here. It is therefore crucial to check that the shock term does not destroy the ESs around the quartic point.
In order to verify this, we launched an input pulse $\psi_{\rm in}=N\sech{\tau}$ in the simulation of Eq. ({\ref{eq2}), and propagate for several dispersion lengths, see Fig. \ref{fig2}, in absence [Fig. \ref{fig2}(a)] and in presence [Fig. \ref{fig2}(b)] of the shock operator, for the parameters indicated in the caption. It is clear by comparing these two figures that the shock term has the effect modifying the velocity of each individual ESs, without affecting their shape or stability during the propagation, and actually allowing the generation of more solitons than what occurs in the case of $s=0$. In order to verify this fact, we compare the approximate analytical non-oscillating solutions of Eq. (\ref{eq2}), for $s=0$, obtained via variational methods, with the actual solitons produced in the simulations (with $s\neq 0$). Such solutions are given by \cite{buryak95}: $\psi(\xi,\tau)=f(\tau)e^{iq\xi}$, $f=a(q){\rm sech}^{2}[k(q)\tau]$, with $a(q)$ and $k(q)$ being the two functions
\eq{var1}{a(q)\equiv\left[\frac{7}{3}k^{2}(q)+\frac{80}{3}|\delta_{4}|k^{4}(q)\right]^{1/2},}
\eq{var2}{k(q)\equiv\left[\frac{\sqrt{9+400|\delta_{4}|q}-3}{80|\delta_{4}|}\right]^{1/2}.} The dashed line in Fig. \ref{fig2}(c) shows the relation between $k(q)$ and $a(q)$ given by Eqs. (\ref{var1}-\ref{var2}), while the dots indicate the parameters of the solitons extracted with a fit from the simulation given in Fig. \ref{fig2}(b) in the ($a$,$k$) space. One can see that the dots are very well predicted by the dashed line, indicating the formation of true ESs in the simulations.

\begin{figure}[htb]
\centering \includegraphics[width=0.5\textwidth]{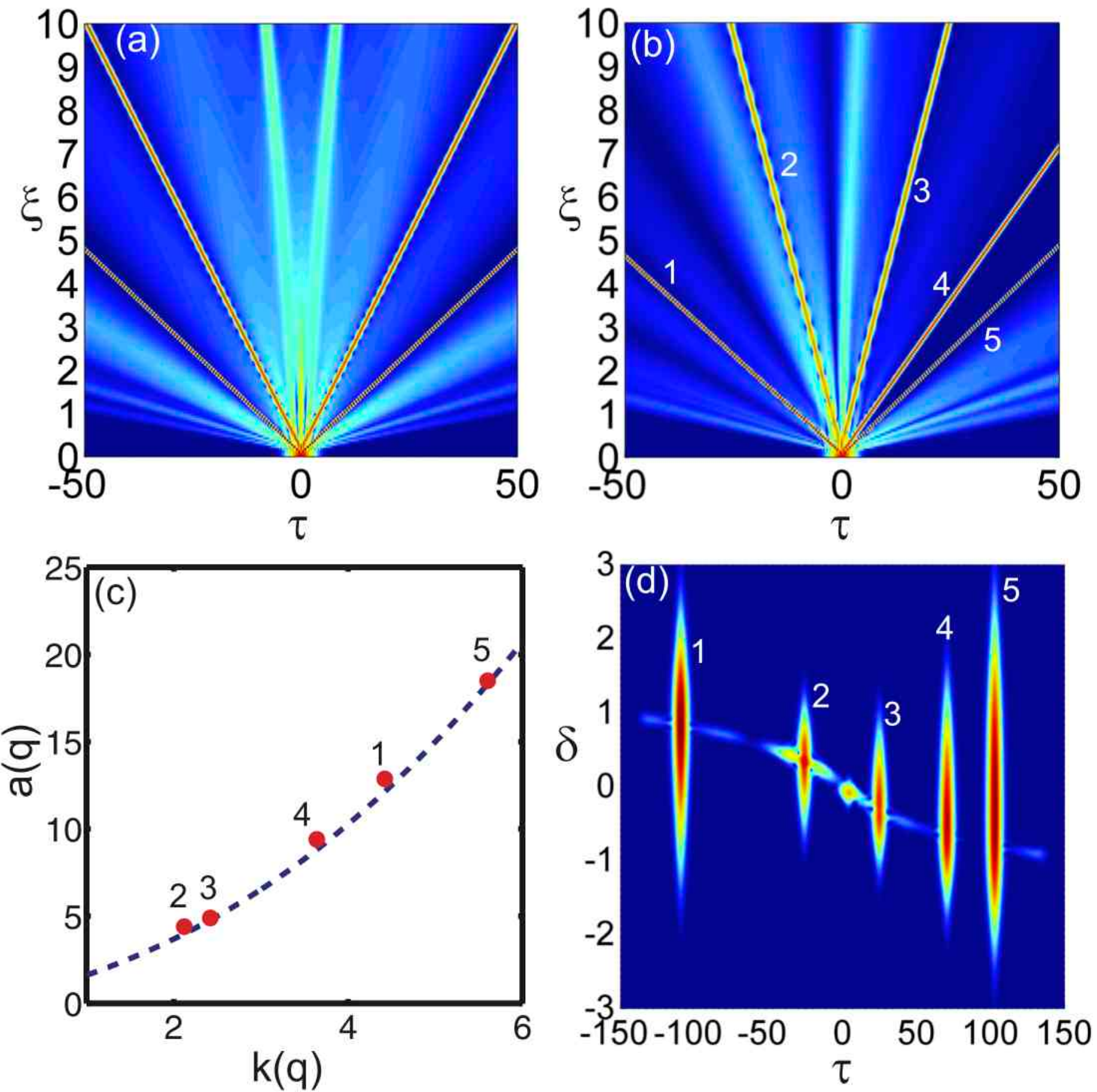}
  \caption{\small{(Color online) (a) Propagation of an input pulse $\psi_{\rm in}=N\sech{\tau}$, with $N=10$ and $\delta_{4}=-0.01$, and no shock term ($s=0$). Due to the non-integrability of Eq. (\ref{eq2}), which induces the presence of a solitonic friction, the pulse splits into several fundamental embedded solitons, approximated by Eqs. (\ref{var1}-\ref{var2}). (b) Same as in (a), but in presence of shock term ($s=0.03$). The shock term induces an asymmetric splitting of the solitons, which all decrease their velocity with respect to the case of (a). (c) Position of the parameters ($a$,$k$) of each solitons formed in (b) after a propagation distance $\xi=10$, on the predicted curve calculated by using Eqs. (\ref{var1}-\ref{var2}). It is evident that all the localized structures formed [which are numbered in (b,c,d)] are embedded solitons. (d) XFROG spectrogram of (b).}}
  \label{fig2}
\end{figure}

Finally, in Fig. \ref{fig3} we show the propagation of a $t_{0}=40$ fs pulse in the structure of Fig. \ref{fig1}(a), for $\lambda_{0}=2.6$ $\mu$m, with other parameters reported in the caption. The formation of a frequency comb extending from approximately $2$ to $3$ $\mu$m is the consequence of the strong spectral interference between many ESs forming near the quartic point of the GVD, which are unable to separate in the frequency domain due to the absence of the Raman effect (and thus of any deceleration that can act on the solitons, followed by a continuous self-frequency shift, \cite{mitschke}) for the chosen quasi-TM input polarization.

\begin{figure}[htb]
\centering \includegraphics[width=0.4\textwidth]{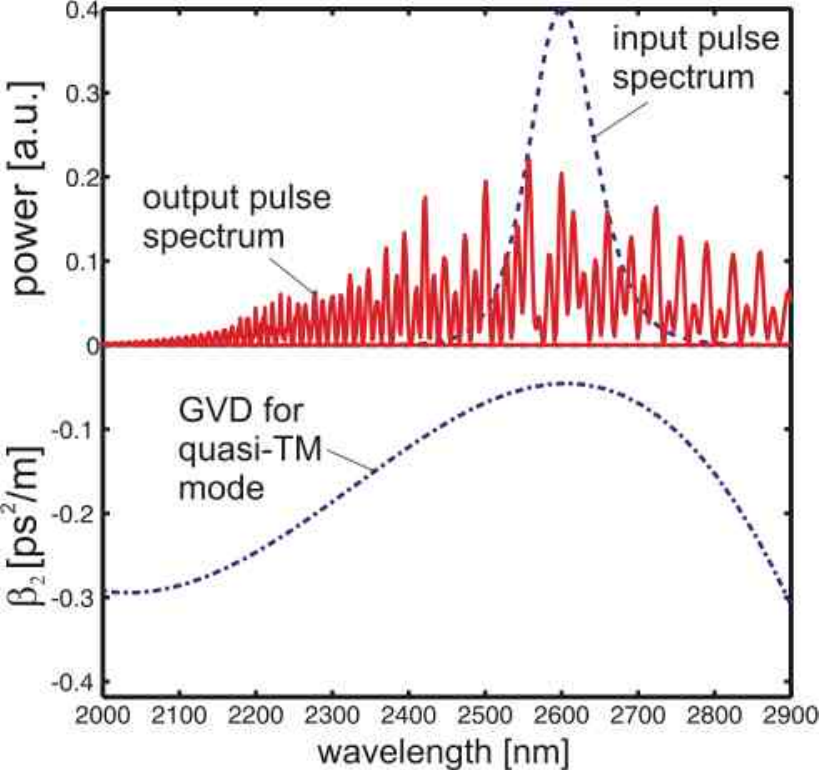}
  \caption{\small{(Color online) Formation of an ESs-induced frequency comb. The bottom part shows the GVD (dashed-dotted line) of the quasi-TM mode vs wavelength for the structure shown in Fig. \ref{fig1}(a). The top part shows the input pulse spectrum (dotted line) and the output spectrum (solid red line) after a propagation of $\xi=10$. Input pulse duration is $t_{0}=40$ fs, input wavelength is $\lambda_{0}=2.6$ $\mu$m, at which $\beta_{2}\simeq -0.05$ ps$^{2}$/m and $\beta_{4}\simeq -2.5\times 10^{-5}$ ps$^{4}$/m. The nonlinear coefficient for the structure of Fig. \ref{fig1}(a) is calculated by using the vector model of Ref. \cite{afshar}, and is $\gamma\simeq 42$ m$^{-1}$W$^{-1}$. The corresponding fundamental power is $P_{0}=0.74$ W. Input pulse shape is $N\sech{t/t_{0}}$, with soliton order $N=5$.}}
  \label{fig3}
\end{figure}


In conclusion, we have demonstrated analytically and numerically the possibility of the observation of localized embedded solitons in specially designed silicon-based slot waveguides. The absence of the Raman effect for TM-polarized pulses, together with specific structures that allow a quartic GVD curve to be realized far from the two-photon absorption region of silicon, permits the generation of many non-radiative ESs and the formation of a spectrally localized frequency comb around the quartic point of the GVD.

This research was funded by the German Max Planck Society for the Advancement of Science (MPG).

\pagebreak

\section*{Informational Fourth Page}

\end{document}